\documentclass[conference]{IEEEtran}
\IEEEoverridecommandlockouts
\usepackage{cite}
\usepackage{amsmath,amssymb,amsfonts}
\usepackage{algorithm}
\usepackage{algorithmic}
\usepackage{graphicx}
\usepackage{textcomp}
\usepackage{xcolor}
\usepackage{dsfont}
\newcommand{\pnext}{.\textit{nx}}
\newcommand{\pprev}{.\textit{pr}}
\newcommand{\msf}{\mathsf}
\usepackage{mathtools}

\newtheorem{remark}{Remark}

\newtheorem{proposition}{Proposition}
\newtheorem{theorem}{Theorem}
\input{widebar_hack.tex}
\def\BibTeX{{\rm B\kern-.05em{\sc i\kern-.025em b}\kern-.08em
    T\kern-.1667em\lower.7ex\hbox{E}\kern-.125emX}}
\begin{document}

\title{{Supporting Passive Users in mmWave Networks}\\
\thanks{The research carried out at UCLA was supported in part by the U.S. National Science Foundation (NSF) awards \mbox{ECCS-2229560} and \mbox{CNS-2146838}.
The work of M. Cardone was supported in part by the NSF under Grants \mbox{CCF-2045237} and \mbox{CNS-2146838}.}
}

\author{\IEEEauthorblockN{Mine Gokce Dogan}
\IEEEauthorblockA{\textit{University of California, Los Angeles} \\
Los Angeles, CA 90095, USA \\
minedogan96@g.ucla.edu}
\and
\IEEEauthorblockN{Martina Cardone}
\IEEEauthorblockA{\textit{University of Minnesota}\\
Minneapolis, MN 55455, USA \\
mcardone@umn.edu}
\and
\IEEEauthorblockN{Christina Fragouli}
\IEEEauthorblockA{\textit{University of California, Los Angeles}\\
Los Angeles, CA 90095, USA \\
christina.fragouli@ucla.edu}
}

\maketitle

\begin{abstract}
The interference from active to passive users is a well-recognized challenge in millimeter-wave (mmWave) communications. 
We propose a method that enables to limit the interference on passive users (whose presence may not be detected since they do not transmit) with a small penalty to the throughput of active users. 
Our approach abstracts away (in a simple, yet informative way) the physical layer component and it leverages
the directivity of mmWave links and the available network path diversity.  We provide linear programming formulations, lower bounds on active users rates, numerical evaluations, and we establish a connection with the problem of (information theoretically) secure communication over mmWave networks.  
\end{abstract}


\section{Introduction}
The fact that active users can destructively interfere with the operation of passive users is a well-recognized challenge in millimeter-wave (mmWave) communications. MmWave infrastructure is increasingly deployed to support a large variety of active user applications, such as virtual reality communications, wearable devices, vehicular networks, and 5G communication systems~\cite{Chittimoju_2021,Wang18,Uwaechia20,Rangan}. 
However, the same spectrum is shared by a number of passive users (i.e., users that do not transmit and can be significantly impacted) such as the Global Positioning System (GPS), passive remote sensing systems, and satellites that study Earth exploration, weather monitoring, and radio-astronomy~\cite{Priebe12, Polese21, Polese22,Bosso21,Eichen19}. In this paper, we propose and evaluate an approach that aims to provide suitable guidelines on how to support a resilient coexistence between passive and active users over mmWave networks.

Supporting the coexistence of passive and active users is a challenging task, for good reasons:  by definition, passive users do not transmit and hence, their presence might not be detected. Moreover, they may be mobile and intermittent, changing their location and their time periods of operation. The question we ask in this paper is, can we guarantee a certain amount of interference-free operation to passive users, while not significantly impacting the experienced performance (communication rates) of the active users?

Our main observation is that, perhaps this is possible over mmWave networks that provide sufficient path diversity. In particular, we propose to constrain each active user to only transmit for up to a desired fraction of time $\theta$ over each link. Due to the directivity of mmWave communications, this translates to that, with very high probability, each passive user will enjoy interference-free operation for a $(1-\theta)$ time fraction. It is not difficult to see that if the capacity of a mmWave network is $\mathsf{C}$, then we can certainly achieve the rate $\theta \mathsf{C}$ with this operation. The interesting part is that, provided that there exists sufficient path diversity over the network, it may  be possible to achieve much higher rates by appropriately designing scheduling and routing schemes.  For instance, our numerical evaluations indicate that for $\theta=0.2$ over randomly generated networks, we can almost always achieve $85\%$ of the unrestricted (oblivious to passive users) network capacity.

Technically, we build on the so-called \hbox{1-2-1} network model that offers a simple yet informative model for mmWave networks~\cite{ezzeldin,EzzeldinISIT2019,EzzeldinISIT2019Multicast}. The model abstracts away the physical layer component and it focuses on capturing an inherent and dominant characteristic of mmWave communications, that is, directivity:
mmWave requires beamforming with narrow beams to compensate for the high path loss incurred by isotropic transmissions. 
Because of this, both the mmWave transmitter and receiver employ antenna arrays to electronically steer and direct their beams towards each other.
This will activate the link between them for communication, which was termed as 1-2-1 link
by the authors in~\cite{ezzeldin}.
In particular,~\cite{ezzeldin} proved that the capacity of a Gaussian noise \hbox{1-2-1} network can be approximated to within a constant (i.e., which only depends on the number of network nodes) additive gap and its optimal beam schedule can be computed in polynomial-time. We leverage the results in~\cite{ezzeldin} to develop efficient transmission and scheduling mechanisms that offer suitable guidelines on how to support the coexistence with passive users in mmWave networks. We analyze the impact of passive users on the approximate network capacity and provide guarantees on the achieved rate. Our main contributions are as follows:
\begin{itemize}
\item For arbitrary mmWave networks, we formulate the problem of finding the maximum
rate achieved while limiting the interference at every node as a Linear Program (LP) and show that it efficiently (i.e., in polynomial time in the network size) finds a beam schedule that supports passive users to desired thresholds.
\item For arbitrary mmWave networks, we derive lower bounds on the active user rates.  We also derive lower bounds on the necessary and sufficient number of paths that achieve a target rate while supporting passive users in arbitrary mmWave networks with unit capacity links. We provide numerical evaluations over randomly generated networks for unequal link capacities.
\item We identify a connection between the passive user problem and the (information theoretically) secure communication problem. For arbitrary mmWave networks with unit capacity links, we prove a reduction between these two problems and provide guarantees on the achieved rates in both problems.
\end{itemize}
\noindent \textbf{Related Work.} 
A multitude of works in the literature study scheduling and routing in wireless networks by exploring multi-path diversity~\cite{Zhou12,Rocher2009,Shillingford2008,YeungBook}. However, these works do not consider passive users and thus, their proposed approaches do not provide interference mitigation between passive and active users. Several works study passive users in passive remote sensing systems and for satellites that study Earth exploration and radio-astronomy~\cite{Reindl2008,Gamba13,Kuypers2006,Alagoz2007}. They study the characterization of these wireless systems and propose routing algorithms or interference mitigation techniques to support passive users. However, these works focus on traditional wireless networks and they do not consider the scheduling constraints of mmWave communications. There exist studies that address path selection and rate allocation problems in \hbox{multi-hop} mmWave networks~\cite{Garcia15,Hu17,Sahoo17,Yan18,Vu19}. However, these works do not consider passive users.
Closer to our work, there exist studies that focus on passive users in directional communication networks, such as mmWave and Terahertz (THz) networks. These works study possible interference scenarios and propose interference mitigation techniques such as employing highly directive and electrically
steerable antennas~\cite{Priebe12}, using spread spectrum techniques~\cite{Bosso21}, or sharing the spectrum between active and passive users~\cite{Polese21,Polese22,Eichen19}. 
However, these works do not propose routing algorithms for path selection and rate allocation, and they do not provide information-theoretical guarantees on the active user rates. Differently, in our work we leverage the directivity of mmWave links and path diversity to develop efficient scheduling mechanisms that support passive users, and derive theoretical guarantees 
on the active user rates.

\noindent \textbf{Paper Organization.} Section~\ref{section_model} provides background on the \hbox{1-2-1} network model.
Section~\ref{section:scheduling} presents the proposed scheduling mechanisms that support passive users, and it provides lower bounds on the active user rates and numerical evaluations. Section~\ref{section:security} introduces a connection between the passive user problem and the (information theoretically) secure communication problem. 
Section~\ref{section_conclusion} concludes the paper.
\section{System Model and Background}\label{section_model}
\textbf{Notation.} 
$[a\!:\!b]$ is the set of integers from $a$ to $b>a$; \hbox{$|\cdot|$} denotes the cardinality for sets and the absolute value for scalars; $\mathds{1}_P$ is the indicator function.

We consider an $N$-relay Gaussian noise Full-Duplex (FD) \hbox{1-2-1} network model where $N$ relays assist the communication between the source node (node $0$) and the destination node \hbox{(node $N+1$)}. At any particular time, each relay
can simultaneously transmit and receive by using a single transmit beam and a single receive beam. Thus, at any particular instance, a relay node can transmit to at most one node and it can receive from at most one node. The source (respectively, destination) can transmit to (respectively, receive from) $M$ other nodes i.e., on $M$ outgoing links (respectively, on $M$ incoming links), simultaneously. Formally, in a Gaussian FD 1-2-1 network, the received signal at node $j \in [1\!:\!N{+}1]$ 
can be written as,
\begin{equation}\label{eq:channel_model}
Y_j = \sum_{i\in[0{:}N]\backslash \{j\}} h_{ji} \mathds{1}_{\{i \in S_{j,r},j \in S_{i,t}\}}X_i + Z_j,
\end{equation}
where: (i) $X_i$ is the channel input at node $i \in [0\!:\!N]$ with power constraint $\mathbb{E}\left[|X_i|^
2\right] \leq P$;
(ii) $h_{ji} \in \mathbb{C}$ is the complex channel coefficient\footnote{The channel coefficients are assumed to be constant for the whole transmission duration and known by the
network.} from node $i$ to node $j$;
(iii) $S_{i,t}$ and $S_{j,r}$ represent the node(s) towards which node $i$ is beamforming its transmissions and the node(s) towards which node $j$ is pointing its receive beam(s);
and (iv) $Z_j$ indicates the additive white Gaussian noise at node $j$; noises across the network are independent and identically distributed as $\mathcal{CN}(0, 1)$.

\begin{remark}
Although \hbox{1-2-1} networks capture
the essence of mmWave communications and enable to derive useful insights on \hbox{near-optimal} information flow algorithms, the model makes a number of simplifying assumptions that include: 
1) 
not considering the overhead of channel knowledge and of \hbox{beam-steering}\footnote{Following beam alignment, the channel time variations are
reduced significantly~\cite{Song18} and hence, the channel state changes much slower than the rate of communication.},
and 2) assuming no interference among communication links (a reasonable assumption for relays spaced further apart than the beam width). 
However, in~\cite{Ezzeldin20} the authors relaxed
this last assumption and considered networks where nodes are equipped with
imperfect beams that have \hbox{side-lobe} leakage. They showed that even with imperfect \hbox{side-lobe}
attenuation, the \hbox{1-2-1} model is a viable approximation when certain operational conditions on the beamforming pattern are satisfied. Thus, for networks satisfying the conditions
introduced in~\cite{Ezzeldin20}, the results naturally
extend.
\end{remark}

We next discuss some known
capacity results for Gaussian FD \hbox{1-2-1} networks for the case $M=1$.

\noindent
\textbf{Capacity of Gaussian FD \hbox{1-2-1} networks.} In~\cite{ezzeldin}, it was shown that the memoryless channel model in~\eqref{eq:channel_model} allows to upper bound the channel capacity using the \hbox{information-theoretic} \hbox{cut-set} upper bound. The authors showed that the unicast capacity of an $N$-relay Gaussian FD \hbox{1-2-1} network can be approximated to within an additive gap that only depends on the number of nodes in the network. In particular, the following LP
was proposed to compute the unicast approximate capacity and its optimal schedule,
\begin{align}\label{capacity_paths}
\begin{array}{llll}
&\ {\rm{P1:}}\  \displaystyle \widebar{\msf{C}} = \underset{x_p, p \in \mathcal{P}}{\max} \displaystyle\sum_{p \in \mathcal{P}} x_p \mathsf{C}_p   & & \\
& ({\rm P1}a) \ x_p \geq 0, & \forall p \!\in\! \mathcal{P}, &  \\
& ({\rm P1}b) \ \displaystyle\sum_{p \in \mathcal{P}_i}  x_p f^p_{p\pnext(i), i} \!\leq\! 1, & \forall i \! \in \! [0\!:\!N], & \\
& ({\rm P1}c) \ \displaystyle\sum_{p \in \mathcal{P}_i} x_p f^p_{i, p\pprev(i)} \!\leq\! 1, & \forall i \! \in \! [1\!:\!N\!+\!1],  &
\end{array}
\end{align}
where: (i) $\widebar{\msf C}$ is the approximate capacity; (ii) $\mathcal{P}$ is the collection of all paths connecting the source to the destination; (iii) $\mathsf{C}_p$ is the capacity of path $p$; (iv) $\mathcal{P}_i \subseteq \mathcal{P}$ is the set of paths that pass through node $i$ where $i \in [0\!:\!N{+}1]$; (v) $p\pnext(i)$ (respectively, $p\pprev(i)$) is the node that follows (respectively, precedes) node $i$ in path $p$; (vi) 
$x_p$ is the fraction of time that path $p$ is used; and (vii) $f^p_{j,i}$ is the optimal activation time for the link of capacity $\ell_{ji}$ when path $p$ is operated, i.e.,
 $   f^p_{j,i} = {\mathsf{C}}_p/\ell_{ji}.$
Here, $\ell_{ji}$ denotes the capacity of the link going from node $i$ to node $j$ where $(i,j) \in  [0\!:\!N]\times[1\!:\!N{+}1]$.

Although the number of variables in 
the LP $\rm{P1}$ (particularly, the number of paths) can be exponential in the number of nodes, this LP
can indeed be solved in polynomial time through the following equivalent LP
as proved in~\cite{ezzeldin}. We refer readers to~\cite{ezzeldin} for a more detailed description.
\begin{align}\label{capacity_lp}
& \ \rm{P2:}\ \widebar{\msf{C}} = \displaystyle\max\limits_{\lambda,F} \hspace{-0.02in}\sum_{j=0}^N F_{(N+1)j}, & & \nonumber \\
&({\rm P2}a) \ 0 \leq F_{ji} \leq \lambda_{ji}\ell_{ji},& \forall (i,j) {\in} [0\!:\!N] \!\times\! [1{:}N{+}1], & \nonumber\\
&({\rm P2}b) \hspace{-0.2in}\displaystyle \sum_{\substack{j \in [1:N{+}1]\backslash \{i\}}} \hspace{-0.27in} F_{ji} = \!\!\! \displaystyle \sum_{\substack{k \in [0{:}N]\backslash \{i\}}} \hspace{-0.21in} F_{ik}, & \forall i \in [1{:}N], & \nonumber\\
&({\rm P2}c) \! \! \! \! \displaystyle \sum_{\substack{j \in [1:N{+}1]\backslash\{i\}}} \! \! \! \! \!\!\hspace{-0.12in} \lambda_{ji} \leq 1, & \forall i \in [0{:}N], & \\
&({\rm P2}d) \! \! \! \! \displaystyle \sum_{\substack{i \in [0:N]\backslash\{j\}}} \! \! \! \! \hspace{-0.1in} \lambda_{ji} \leq 1, & \forall j \in [1{:}N+1], & \nonumber\\
&({\rm P2}e) \ \lambda_{ji} \geq 0,  &\forall (i,j)  {\in} [0\!:\!N] \!\times\! [1\!:\!N{+}1] \nonumber,
\end{align}
where: (i) $F_{ji}$ denotes the amount of information flowing from node $i$ to node $j$; and (ii) $\lambda_{ji}$ denotes the fraction of time for which the link going from node $i$ to node $j$ is active, based on the schedule used to align the antenna beams in the network. 

\begin{remark}
In the LP $\rm{P1}$ (equivalently, in the LP $\rm{P2}$), the beam scheduling allows to share traffic across multiple paths, both over space and time without considering the interference on passive users. Our goal is to identify which paths to use and how to schedule them so as to limit passive user interference with a small penalty on the throughput of the active users.
\end{remark}

\section{Scheduling Mechanisms for Passive Users}\label{section:scheduling}
In this section, we aim to build scheduling mechanisms that support the coexistence of active and passive users over arbitrary mmWave networks.  

We assume that every node $i\in[0\!:\!N]$ in the network is allowed to actively transmit over each network link $(i,j)$ for at most $0\leq \theta_{ji} \leq 1$ fraction of time where $j \in [1\!:\!N{+}1]$.  That is, each link 
can be used for at most a certain fraction of time.  These thresholds on the link activation times can be selected depending on the application of interest, side knowledge on the existence of passive users and their requirements, or, using our analysis over a specific network, they can be maximized subject to the constraint that the active user rates are not significantly affected. To account for these thresholds, we
add the following constraint to the LP $\rm{P2}$ in~\eqref{capacity_lp}, which ensures that the activation time of each link is below a certain threshold, i.e., the optimal beam schedule can limit the interference on the passive users,
\begin{equation}\label{passive_constraint}
\lambda_{ji} \leq \theta_{ji}, \quad \forall (i,j)  {\in} [0\!:\!N] \!\times\! [1\!:\!N{+}1],
\end{equation}
where $\theta_{ji}$ denotes the threshold on the activation time $\lambda_{ji}$ of the link going from node $i$ to node $j$. 

\begin{remark}
The LP $\rm{P2}$ in~\eqref{capacity_lp} with the 
constraint in~\eqref{passive_constraint} has a polynomial number of variables and constraints; hence, an off-the-shelf LP solver can solve it 
in polynomial time. By taking the same steps as in~\cite{ezzeldin}, we can show that there is a polynomial-time mapping between the optimal solution of the LP $\rm{P2}$ in~\eqref{capacity_lp} with the additional constraint in~\eqref{passive_constraint} and a feasible beam schedule. Thus, the LP $\rm{P2}$ in~\eqref{capacity_lp} with the constraint in~\eqref{passive_constraint} efficiently finds a feasible beam schedule that supports passive users while maximizing the rate of active users.
\end{remark}

\subsection{Lower Bounds on the Passive Capacity}
Although the LP $\rm{P2}$ in~\eqref{capacity_lp} with the constraint in~\eqref{passive_constraint} can be solved efficiently, it does not provide a \hbox{closed-form} expression for the \hbox{1-2-1} passive capacity $\mathsf{C}$, i.e., the supremum of all rates achieved under the passive users constraint in~\eqref{passive_constraint}. With the goal of further investigating the  \hbox{1-2-1} passive capacity, we here provide a few \hbox{closed-form} lower bounds on it. We start by noting that a simple lower bound on $\mathsf{C}$ is given by

 \begin{equation}\label{eq:lb_passive_capacity}
\mathsf{C} \geq \hat{\theta}\widebar{\mathsf{C}}, 
\end{equation}
where: (i) $\widebar{\mathsf{C}}$ denotes the approximate capacity found by solving the LP $\rm{P2}$ in~\eqref{capacity_lp} without the constraint in~\eqref{passive_constraint}; and (ii) $\hat{\theta}{=}\min_{\theta \in \Theta}\theta$, where $\Theta {=} \left\{\theta_{ji}\mid (i,j)  {\in} [0\!:\!N] \!\times\! [1\!:\!N{+}1]\right\}$ denotes the set of threshold values on the activation times of the network links. 
Indeed, we can achieve $\hat{\theta}\widebar{\mathsf{C}}$ by multiplying every link activation time and link flow in the optimal solution of the LP $\rm{P2}$ in~\eqref{capacity_lp} (without the constraint in~\eqref{passive_constraint}) by $\hat{\theta}$. Since this solution satisfies the constraints in the LP $\rm{P2}$ and the passive users constraint in~\eqref{passive_constraint}, it is a lower bound on $\mathsf{C}$. 

We now use the optimal solution of the LP $\rm{P2}$ in~\eqref{capacity_lp} (without the constraint in~\eqref{passive_constraint}) that is denoted by $\left\{\lambda^\star,~F^\star\right\}$ to derive a tighter bound on the \hbox{1-2-1} passive capacity than the one in~\eqref{eq:lb_passive_capacity}. We let $\tilde{\Theta}$ denote the set of threshold values on the activation times of the links that are activated by the optimal solution $\left\{\lambda^\star,~F^\star\right\}$, i.e., $\tilde{\Theta} {=} \left\{\theta_{ji}\mid \lambda^\star_{ji}>0,~(i,j)  {\in} [0\!:\!N] \!\times\! [1\!:\!N{+}1]\right\}$. We also let $\Lambda$ denote the set of activation times of the links that are activated by $\left\{\lambda^\star,~F^\star\right\}$, i.e., $\Lambda{=} \left\{\lambda^\star_{ji}\mid (i,j)  {\in} [0\!:\!N] \!\times\! [1\!:\!N{+}1]\right\}$.
Then, the following lower bound on $\mathsf{C}$ holds,
\begin{equation}\label{eq:lb1_passive_capacity}
\mathsf{C} \geq \frac{\tilde{\theta}}{\tilde{\lambda}}\widebar{\mathsf{C}},
\end{equation}
where $\tilde{\theta} {=} \min_{\theta \in \tilde{\Theta}} \theta$ and $\tilde{\lambda} {=} \max_{\lambda \in \Lambda} \lambda$. We can achieve the bound in~\eqref{eq:lb1_passive_capacity} by multiplying every element in $\left\{\lambda^\star,~F^\star\right\}$ by $\tilde{\theta}/\tilde{\lambda}$.
This solution satisfies the constraints in the LP $\rm{P2}$ in~\eqref{capacity_lp} and the passive users constraint in~\eqref{passive_constraint}. Since $\tilde{\lambda} \leq 1$ and $\hat{\theta} \leq \tilde{\theta}$, it follows that~\eqref{eq:lb1_passive_capacity} is a tighter lower bound on $\mathsf{C}$ than~\eqref{eq:lb_passive_capacity}. We also note that~\eqref{eq:lb1_passive_capacity} is a lower bound on $\mathsf{C}$ only if $\tilde{\theta} \leq \tilde{\lambda}$. Otherwise, the passive capacity $\mathsf{C}$ is equal to $\widebar{\mathsf{C}}$. 

Since, as proved in \cite{ezzeldin}, the LP $\rm{P1}$ in~\eqref{capacity_paths} and the LP $\rm{P2}$ in~\eqref{capacity_lp} are equivalent LPs, we can use an optimal solution of the LP $\rm{P1}$ to derive a tighter bound on $\mathsf{C}$. We let $\mathcal{P}^\star$ denote the set of active paths in the optimal solution of the LP $\rm{P1}$ (without considering passive users) and $x^\star_p$ denote the optimal activation time of the path $p \in \mathcal{P}^\star$. We also let $\mathcal{E}_p$ denote the set of links of path $p \in \mathcal{P}^\star$.
The following proposition (proof in Appendix~\ref{appendix:lb_passive_capacity}) presents a tighter lower bound on $\mathsf{C}$ than~\eqref{eq:lb_passive_capacity} and~\eqref{eq:lb1_passive_capacity} by leveraging the paths in $\mathcal{P}^\star$. 
\begin{proposition}\label{lemma:lb_passive_capacity}
For an $N$-relay Gaussian FD \hbox{1-2-1} network with an arbitrary topology, the \hbox{1-2-1} passive capacity $\mathsf{C}$ can be lower bounded as follows, 
\begin{equation}\label{eq:lb2_passive_capacity}
\mathsf{C} \geq \sum_{p \in \mathcal{P}^\star} \min\left(x^\star_p,\frac{\tilde{\theta}_p x^\star_p}{\tilde{\lambda}_p} \right)\mathsf{C}_p,
\end{equation}
where $\tilde{\theta}_p {=} \min_{(i,j)\in\mathcal{E}_p} \theta_{ji}$ and $\tilde{\lambda}_p {=} \max_{(i,j)\in\mathcal{E}_p} \lambda^\star_{ji}$.
\end{proposition}
\begin{remark}
We note that $\tilde{\theta}_p {=} \min_{(i,j)\in\mathcal{E}_p} \theta_{ji} \geq \tilde{\theta}$ and $\tilde{\lambda}_p {=} \max_{(i,j)\in\mathcal{E}_p} \lambda^\star_{ji} \leq \tilde{\lambda}$. This readily implies that
$\tilde{\theta}/\tilde{\lambda}$ in~\eqref{eq:lb1_passive_capacity} is smaller than or equal to $\tilde{\theta}_p/\tilde{\lambda}_p$ $\forall p \in \mathcal{P}^\star$. Thus, the bound in~\eqref{eq:lb2_passive_capacity} is a tighter bound than the one in~\eqref{eq:lb1_passive_capacity}.
\end{remark}

Proposition~\ref{lemma:lb_passive_capacity} shows that the paths activated by an optimal solution of the LP $\rm{P1}$ in~\eqref{capacity_paths} can be leveraged to achieve the lower bound 
in~\eqref{eq:lb2_passive_capacity}. In the example below, we highlight that we can indeed achieve a higher rate than the lower bound in~\eqref{eq:lb2_passive_capacity} by distributing the traffic across a larger number of paths.

\begin{figure}
     \centering
     \includegraphics[width=0.5\columnwidth]{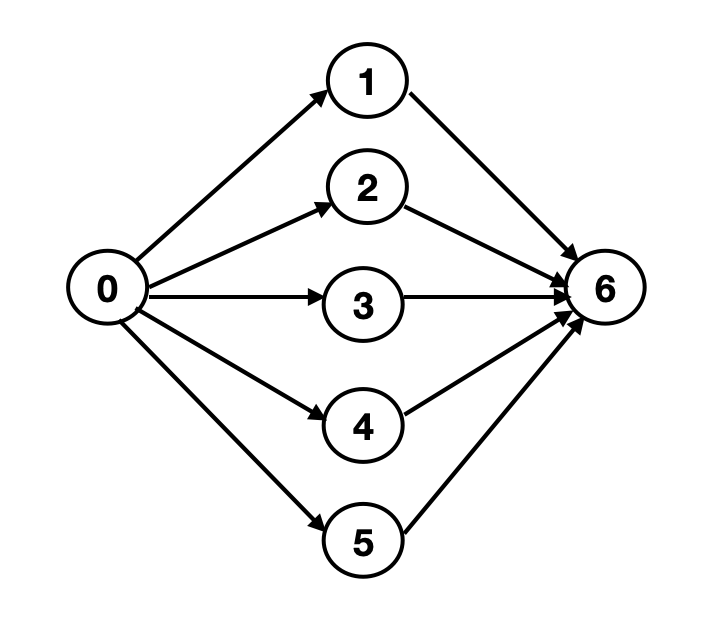}
     \vspace{-0.15in}
     \caption{An mmWave network example with $5$ relay nodes.}
     \label{fig:multipath_example}
     \vspace{-0.05in}
\end{figure}

\noindent \textit{Example 1}. Consider the network with $N=5$ relay nodes in Fig.~\ref{fig:multipath_example}.
There exist $5$ paths connecting the source (node $0$) to the destination (node $6$), namely $p_1: 0 \rightarrow 1 \rightarrow 6,~p_2: 0 \rightarrow 2 \rightarrow 6,~p_3: 0 \rightarrow 3 \rightarrow 6,~ p_4: 0 \rightarrow 4 \rightarrow 6$, and $p_5: 0 \rightarrow 5 \rightarrow 6$. We consider unit capacity links except for the links in $p_1$ for which the link capacities are equal to $2$. We assume $M=1$ and $\theta_{ji} = 0.2$ $\forall (i,j)  {\in} [0\!:\!N] \!\times\! [1\!:\!N{+}1]$. The optimal solution of the LP $\rm{P1}$ in~\eqref{capacity_paths} (without the passive users constraint in~\eqref{passive_constraint}) activates only path $p_1$ to achieve the approximate capacity $\widebar{\mathsf{C}} = 2$. We can reduce the activation time of $p_1$ to $0.2$ in order to satisfy the constraint in~\eqref{passive_constraint}, and this would achieve the rate $0.4$, which is equal to the lower bound in Proposition~\ref{lemma:lb_passive_capacity}. However, if we perform equal time sharing across all of the $5$ paths, each path is activated for $0.2$ fraction of time, and the constraint in~\eqref{passive_constraint} is still satisfied. This solution achieves a rate equal to $1.2$. \hfill $\square$

In Example 1, we showed that we can leverage the path diversity in a network to achieve a higher rate than the lower bound in Proposition~\ref{lemma:lb_passive_capacity}.  A question that naturally arises is: how many paths would be sufficient to achieve a certain target rate while limiting the interference on the passive users? Or, are there any intrinsic properties of the paths that should be leveraged? 

\subsection{Number of Paths for Target Rates}
Here, we provide an answer to the questions above.
Towards this end, we let $H_e$ (respectively, $H_v$) denote the maximum number of \hbox{edge-disjoint} (respectively, \hbox{vertex-disjoint}) paths connecting the source to the destination in the network. The next theorem (proof in Appendix~\ref{appendix:numberOfPaths}) provides lower bounds on $H_e$ and $H_v$ that ensure target rates for the active users.

\begin{theorem}\label{lemma:numberOfPaths}
Consider an $N$-relay Gaussian FD \hbox{1-2-1} network with an arbitrary topology and unit capacity links, and let $\theta$ be the threshold on the activation times of the links in the network. Then, the LP $\rm{P2}$ in~\eqref{capacity_lp} (without the passive users constraint in~\eqref{passive_constraint}) outputs $\bar{\mathsf{C}}$, and
the following holds:

\noindent$\bullet$ For $M=1$: The rate $\theta_c \widebar{\mathsf{C}}$ can be achieved if and only if
\begin{equation}\label{eq:lb1_numOfPaths}
H_e \geq \frac{\theta_c }{\theta} \widebar{\mathsf{C}},
\end{equation}
where $0\leq \theta_c \leq 1$, and $\widebar{\mathsf{C}}=1$.

\noindent$\bullet$ For $M>1$: The rate $\theta_c \widebar{\mathsf{C}}$ can be achieved whenever 
\begin{equation}\label{eq:lb2_numOfPaths}
H_v \geq \frac{\theta_c }{\theta} \widebar{\mathsf{C}},
\end{equation}
where $0\leq \theta_c \leq 1$, and $\widebar{\mathsf{C}} = \min\left(M,H_v\right)$.
\end{theorem}
\begin{remark}
Theorem~\ref{lemma:numberOfPaths} can be directly extended to the 
case in which there exists a threshold $\theta_{ji}$ on the activation time of the link going from node $i$ to node $j$, $\forall (i,j)  {\in} [0\!:\!N] \!\times\! [1\!:\!N{+}1]$. To this end, we can simply replace $\theta$ in~\eqref{eq:lb1_numOfPaths} and~\eqref{eq:lb2_numOfPaths} with the minimum threshold value $\hat{\theta}$ which was defined in~\eqref{eq:lb_passive_capacity}, and find lower bounds on the number of paths to use for target rates.
\end{remark}

\begin{figure}
	\centering
     \includegraphics[width=0.8\columnwidth]{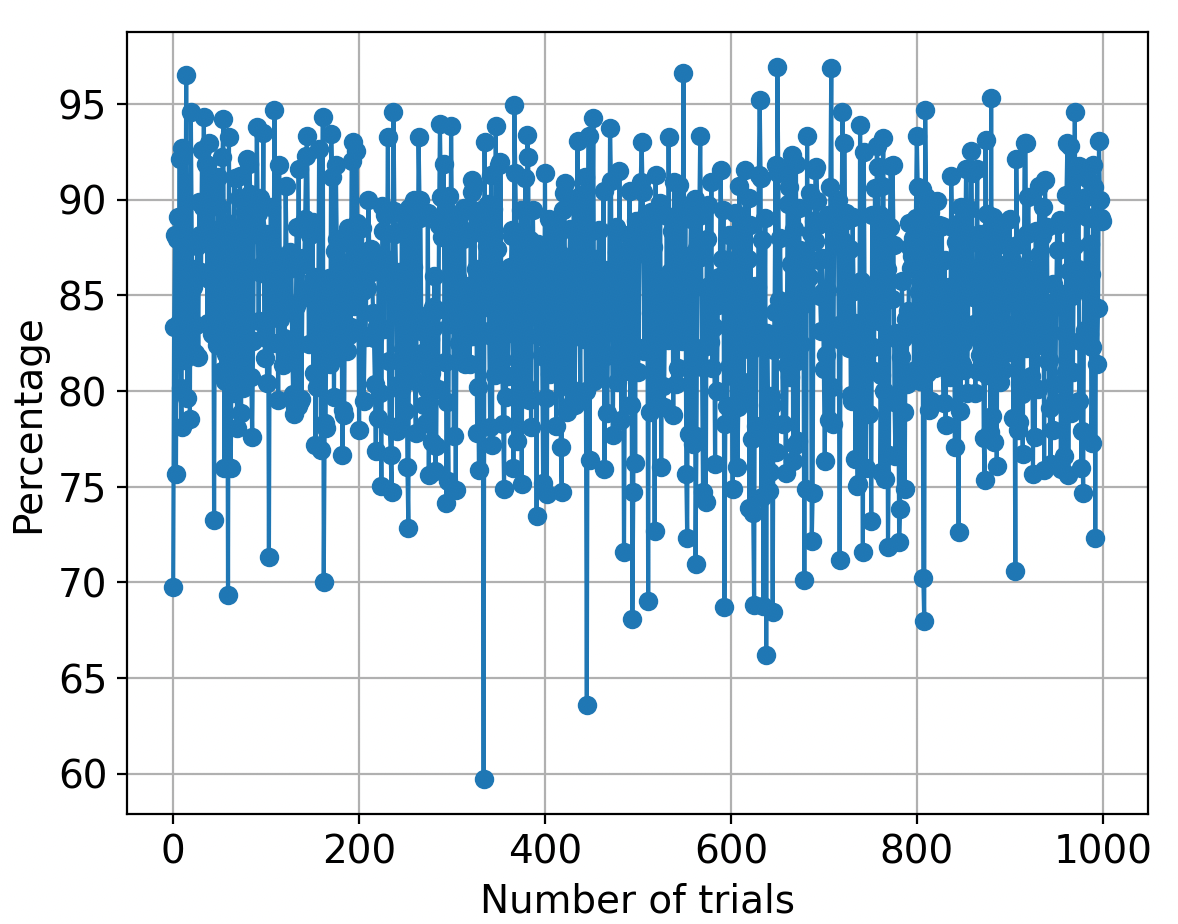}
     \vspace{-0.1in}
     \caption{Achieved percentage of the approximate capacity.}
     \label{percentage_capacity}
     \vspace{-0.1in}
\end{figure}

\begin{figure}
	\centering
     \includegraphics[width=0.8\columnwidth]{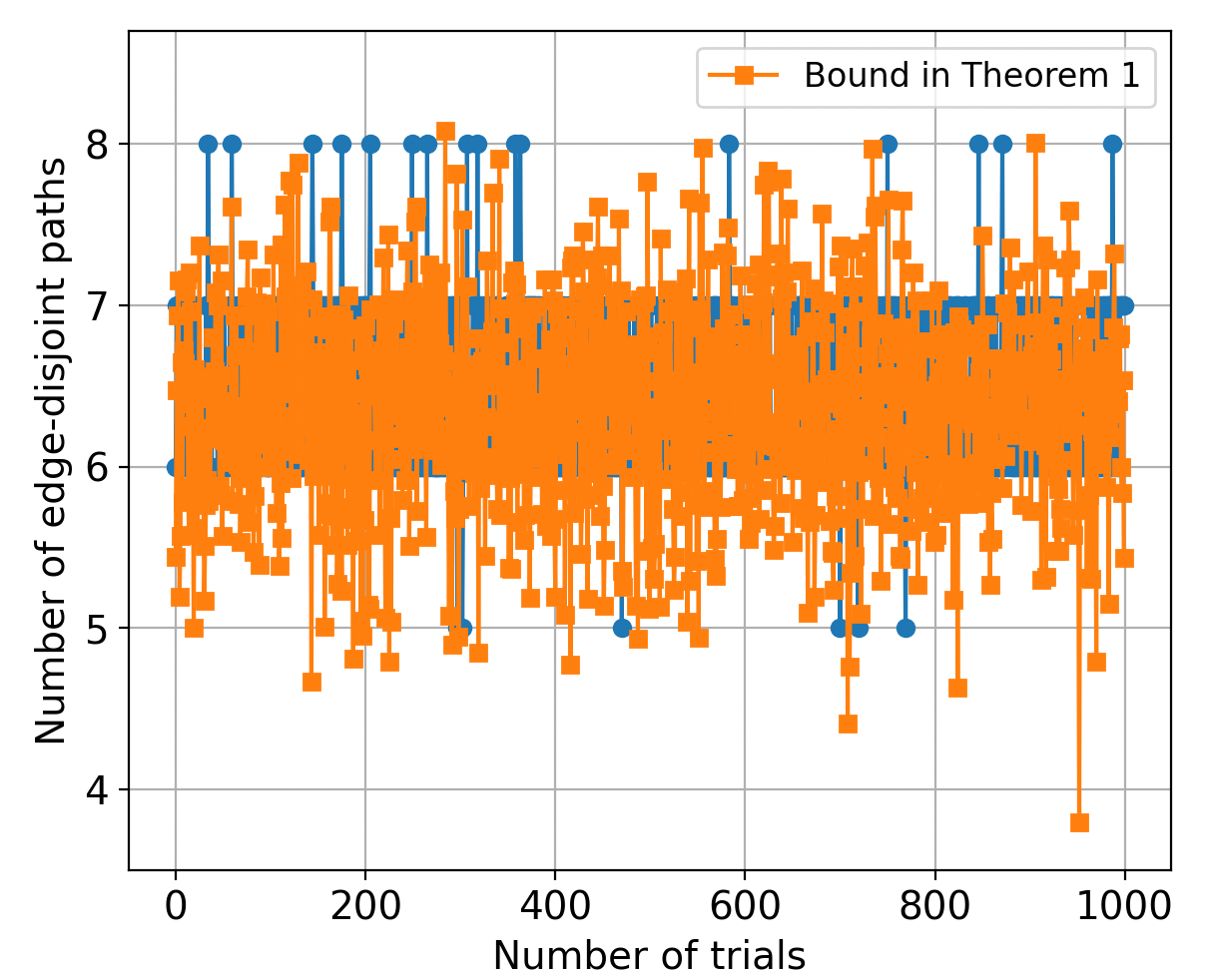}
     \vspace{-0.1in}
     \caption{Maximum number of edge-disjoint paths.}
     \label{edgedisjoint_paths}
     \vspace{-0.1in}
\end{figure}

The lower bounds in Theorem~\ref{lemma:numberOfPaths} were derived for networks with unit capacity links. We next numerically evaluate the \hbox{1-2-1} passive capacity $\mathsf{C}$ and $H_e$ for networks with \textit{unequal} link capacities. Towards this end, we randomly generated a network with $N=10$ relay nodes and performed $1000$ trials over this network. 
In each trial, we generated a different set of link capacities from the Gaussian distribution with mean $1$ and variance $0.1$. We assumed that $M=1$ and we set the threshold on the link activation times equal to $\theta = 0.2$. In Fig.~\ref{percentage_capacity}, we plotted the achieved percentage of the approximate capacity $\widebar{\mathsf{C}}$ over $1000$ trials while satisfying the constraint in~\eqref{passive_constraint}.
From Fig.~\ref{percentage_capacity}, we note that, on average, the passive capacity $\mathsf{C}$ is approximately $85\%$ of $\widebar{\mathsf{C}}$. This shows that, although the activation times of the links in the network can be at most $0.2$, our LP formulation finds a schedule that, in every trial, achieves a rate much higher than the naive lower bound in~\eqref{eq:lb_passive_capacity} equal to $0.2\widebar{\mathsf{C}}$. Thus, we decrease the penalty on the active users throughput. In Fig.~\ref{edgedisjoint_paths}, we show the maximum number of edge-disjoint paths activated by an optimal solution of the LP $\rm{P2}$ in~\eqref{capacity_lp} with the constraint in~\eqref{passive_constraint}. 
Although we consider unequal link capacities, the maximum number of active edge-disjoint paths varies closely around the bound in~\eqref{eq:lb1_numOfPaths} for $\theta_c = 1$. Our evaluation shows that if the link capacities have a small variance, the lower bound in~\eqref{eq:lb1_numOfPaths} still gives a reasonable estimate on the required number of \hbox{edge-disjoint} paths to achieve $\widebar{\mathsf{C}}$ (or the target rate $\theta_c\widebar{\mathsf{C}}$).

\section{Connection to Secure Communication}\label{section:security}
In this section, we present a connection between the passive user problem investigated in Section~\ref{section:scheduling}, and the (information theoretically) secure communication problem. In particular, we
show a reduction between these two problems and provide guarantees on the achieved rates.

In the (information theoretically) secure communication problem over \hbox{1-2-1} networks~\cite{Agarwal18}, an arbitrary \hbox{1-2-1} network with unit capacity links is considered, where a passive eavesdropper wiretaps any $K$ links of the network. 
The authors showed that the source can securely communicate with the destination at high rates 
by leveraging directivity and multipath diversity in mmWave networks. Particularly, the source can vary which paths it operates over time and this is possible thanks to the fact that we may have several possible choices of paths to achieve the unsecure capacity.
Thus, in the secure communication problem over \hbox{1-2-1} networks, the traffic is distributed across multiple paths to achieve a high secure rate. This is similar to the passive user problem where we again distribute the traffic across multiple paths to ensure that the activation time of each link in the network is below a certain threshold.
Thus, we here aim to perform a reduction between the passive user problem and the secure communication problem. Particularly, we leverage 
a \hbox{high-performing} (e.g., sometimes capacity achieving)
scheme for one problem and see what rate it can guarantee for the other.

In the secure communication problem, we consider the case where  a passive eavesdropper can wiretap any $K$ links of the network. 
In the passive user problem, recall that $\theta_{ji}$ is the threshold on the activation time of the link going from node $i$ to node $j$, $\forall (i,j)  {\in} [0\!:\!N] \!\times\! [1\!:\!N{+}1]$ (see~\eqref{passive_constraint}). 
We let $H_e$ denote the maximum number of \hbox{edge-disjoint} paths connecting the source to the destination, 
and we denote the corresponding set of \hbox{edge-disjoint paths} by $p_{[1:H_e]}\subseteq \mathcal{P}$. 
Similarly, we let $H_v$ be the maximum number of \hbox{vertex-disjoint} paths connecting the source to the destination,
and we denote the corresponding set of \hbox{vertex-disjoint} paths by $p_{[1:H_v]}\subseteq \mathcal{P}$. Theorem~\ref{lemma:security_connection1} (proof in Appendix~\ref{appendix:security_connection1}) 
formally presents our results.
\begin{theorem}\label{lemma:security_connection1}
Consider an $N$-relay Gaussian FD 1-2-1 network with an arbitrary topology and unit capacity links. 
Let $H_e$ (respectively, $H_v$) denote the maximum number of \hbox{edge-disjoint} (respectively, \hbox{vertex-disjoint}) paths connecting the source to the destination.
Then the following holds:

\noindent$\bullet$ By using the paths activated by an optimal passive user scheme and the corresponding path activation times, we can guarantee a secure rate $R$ 
such that,
\begin{equation}\label{eq:passive_to_secure}
R = \mathsf{C}-\max_{b \in \mathcal{B}}\sum_{(i,j) \in b}\theta_{ji},
\end{equation}
where $\mathsf{C}$ denotes the \hbox{1-2-1} passive capacity and $\mathcal{B}$ is the set of all combinations of $K$ links.

\noindent$\bullet$ By leveraging the paths activated by a secure communication scheme and the corresponding path activation times, we can guarantee a rate $R$ in the passive user problem such that,
\begin{align}\label{eq:secure_to_passive}
R = 
\left \{
\begin{array}{ll}
\sum_{p \in p_{[1:H_e]}}\min\left(\frac{1}{H_e},\theta_p\right) & \text{if} \ M=1,
\\
\sum_{p \in p_{[1:H_v]}}\min\left(\frac{M}{H_v},\theta_p\right) & \text{if} \ M>1,
\end{array}
\right .
\end{align}
where $\theta_p {=} \min_{(i,j) \in \mathcal{E}_p}\theta_{ji}$.
\end{theorem}

We now show that the connection presented in Theorem~\ref{lemma:security_connection1} is useful as there exist scenarios in which the same set of paths characterize both the passive and secure capacities. Toward this end, we consider the case $M=1$ and for the passive user problem, we assume that $\theta$ is the threshold on the activation times of the links in the network. For $M=1$, the optimal passive user scheme is found by solving the LP $\rm{P2}$ in~\eqref{capacity_lp} with the constraint in~\eqref{passive_constraint}. As we showed in Appendix~\ref{appendix:numberOfPaths}, there exists an optimal solution that activates the maximum number of \hbox{edge-disjoint} paths. In~\cite{Agarwal18}, the optimal secure scheme that achieves the secure capacity also activates the maximum number of \hbox{edge-disjoint} paths. Thus, the set of \hbox{edge-disjoint} paths $p_{[1:H_e]}$ characterize both the passive capacity and the secure capacity. For example, consider the network in Fig.~\ref{fig:multipath_example} with unitary link capacities and $\theta = 0.2$. There exist $H_e=5$ \hbox{edge-disjoint} paths in the network. An optimal solution for the passive user problem activates all five paths such that the activation time of each path is equal to $0.2$ and the passive capacity is $\mathsf{C}=1$. This is equal to the rate $R$ in~\eqref{eq:secure_to_passive}. The optimal secure scheme in~\cite{Agarwal18} also activates all five paths such that the activation time of each path is equal to $0.2$ and the secure capacity is equal to $1-K/H_e = 1-0.2K$. This is equal to the rate $R$ in~\eqref{eq:passive_to_secure}. Thus, there exist scenarios where the lower bounds in Theorem~\ref{lemma:security_connection1} are tight and exactly equal to the capacity. In the general case where each link $(i,j)$ has a different threshold $\theta_{ji}$, the set of \hbox{edge-disjoint} paths $p_{[1:H_e]}$ and equal time sharing across these paths with activation time $1/H_e$ characterize both the passive capacity and the secure capacity if $1/H_e \leq \hat{\theta}$, where $\hat{\theta}$ is defined in~\eqref{eq:lb_passive_capacity}. In the above discussion, we have considered the case $M=1$; however, it is possible to extend the analysis when $M>1$, i.e., also for this case there are scenarios in which the same set of paths characterize both the passive and the secure capacities.

\section{Conclusions}\label{section_conclusion}
We have proposed and evaluated an approach that aims to support resilient coexistence of passive and active users in mmWave networks. 
Our aim was to guarantee a certain amount of interference-free operation to passive users, while not significantly impacting the 
rates of the active users. 
We formulated the problem of finding the maximum rate achieved in arbitrary mmWave networks, while limiting the interference at every node, as an LP. We derived lower bounds on the rates of the active users and on the number of paths that can achieve a target rate, while supporting passive users. 
We numerically evaluated our results, which showcase the effectiveness of our approach, e.g., $85\%$ of the unrestricted (oblivious to passive users) network capacity can be achieved even when
$\theta=0.2$.
Finally, we established a connection between the passive user problem and the problem of (information theoretically) secure communication in mmWave networks.
In particular, we performed a reduction between the two problems,
and we showed that there are scenarios in which the same set of paths can characterize both the passive and the secure capacities.

\bibliographystyle{IEEEtran}
\bibliography{bibliography}
\appendices
\section{Proof of Proposition~\ref{lemma:lb_passive_capacity}}\label{appendix:lb_passive_capacity}
In the proof, we show that there exists a feasible solution for the LP $\rm{P1}$ in~\eqref{capacity_paths} that satisfies the passive users constraint in~\eqref{passive_constraint} and achieves the rate in~\eqref{eq:lb2_passive_capacity}.

Let $\hat{x}_p {=}\min\left(x^\star_p,\tilde{\theta}_p x^\star_p/\tilde{\lambda}_p \right)$, $\forall p \in \mathcal{P}^\star$. The proof shows that activating each path $p \in \mathcal{P}^\star$ with the activation time $\hat{x}_p$ is a feasible solution for the LP $\rm{P1}$ in~\eqref{capacity_paths} and it satisfies the passive users constraint in~\eqref{passive_constraint}. 

By construction, the constraint $(\rm{P1}a)$ is satisfied by $\hat{x}_p$ since $x^\star_p$, $\tilde{\theta}_p$, and $\tilde{\lambda}_p$ are all nonnegative for every $p \in \mathcal{P}^\star$. 
Moreover, $\left\{ x_p^\star, \forall p \in \mathcal{P}^\star \right \}$ is an optimal solution for the LP $\rm{P1}$ and hence, it satisfies the constraints in $(\rm{P1}b)$ and $(\rm{P1}c)$. Since $\hat{x}_p \leq x^\star_p$ for every $p \in \mathcal{P}^\star$, the constraints in $(\rm{P1}b)$ and $(\rm{P1}c)$ are satisfied by $\left\{ \hat{x}_p, \forall p \in \mathcal{P}^\star \right \}$
as well. Thus, $\left\{ \hat{x}_p, \forall p \in \mathcal{P}^\star \right \}$ is a feasible solution for the LP $\rm{P1}$. 

We next show that $\left\{ \hat{x}_p, \forall p \in \mathcal{P}^\star \right \}$ also satisfies the passive users constraint in~\eqref{passive_constraint}. The activation time of the link going from node $i$ to node $j$ can be written in terms of path activation times as $\sum_{p \in \mathcal{P}^\star_{ji}}\hat{x}_p\left(\mathsf{C}_p/\ell_{ji}\right)$, where $\mathcal{P}^\star_{ji}$ denotes the set of paths in $\mathcal{P}^\star$ that pass through the link going from node $i$ to node $j$. Thus, the constraint in~\eqref{passive_constraint} is satisfied if
\begin{equation*}
\sum_{p \in \mathcal{P}^\star_{ji}}\hat{x}_p\frac{\mathsf{C}_p}{\ell_{ji}} \leq \theta_{ji}, \quad \forall (i,j)  {\in} [0\!:\!N] \!\times\! [1\!:\!N{+}1].
\end{equation*}
We note that
\begin{equation*}
\sum_{p \in \mathcal{P}^\star_{ji}}\hat{x}_p\frac{\mathsf{C}_p}{\ell_{ji}} \stackrel{{\rm{(a)}}}{\leq}  \sum_{p \in \mathcal{P}^\star_{ji}}\frac{\tilde{\theta}_p x^\star_p \mathsf{C}_p}{\tilde{\lambda}_p \ell_{ji}} \stackrel{{\rm{(b)}}}{\leq}
\sum_{p \in \mathcal{P}^\star_{ji}}\frac{\theta_{ji}x^\star_p \mathsf{C}_p}{\lambda^\star_{ji}\ell_{ji}},
\end{equation*}
where $\rm{(a)}$ is due to the fact that $\hat{x}_p \leq \left(\tilde{\theta}_px^\star_p/\tilde{\lambda}_p\right)$ for all $p \in \mathcal{P}^\star$, and $\rm{(b)}$ follows from the definitions of  $\tilde{\theta}_p$ and $\tilde{\lambda}_p$ for $p \in \mathcal{P}^\star$, which imply that 
$\left(\tilde{\theta}_p/\tilde{\lambda}_p\right) \leq \left(\theta_{ji}/\lambda^\star_{ji}\right)$, $\forall (i,j) \in \mathcal{E}_p$. 
Thus, from the above we obtain
\begin{equation*}
\sum_{p \in \mathcal{P}^\star_{ji}}\hat{x}_p\frac{\mathsf{C}_p}{\ell_{ji}} \leq \frac{\theta_{ji}}{\lambda^\star_{ji}}\sum_{p \in \mathcal{P}^\star_{ji}}x^\star_p\frac{\mathsf{C}_p}{\ell_{ji}}.
\end{equation*}
Since $\lambda^\star_{ji} {=} \sum_{p \in \mathcal{P}^\star_{ji}}x^\star_p\left(\mathsf{C}_p/\ell_{ji}\right)$, we finally have that
\begin{equation*}
 \sum_{p \in \mathcal{P}^\star_{ji}}\hat{x}_p\frac{\mathsf{C}_p}{\ell_{ji}} \leq \theta_{ji}, \quad \forall (i,j)  {\in} [0\!:\!N] \!\times\! [1\!:\!N{+}1].
\end{equation*}
Thus, the constraint in~\eqref{passive_constraint} is satisfied by $\left\{ \hat{x}_p, \forall p \in \mathcal{P}^\star \right \}$, and the rate achieved with this choice is equal to the right-hand side of~\eqref{eq:lb2_passive_capacity}. This concludes the proof of Proposition~\ref{lemma:lb_passive_capacity}.

\section{Proof of Theorem~\ref{lemma:numberOfPaths}}\label{appendix:numberOfPaths}
In the proof, we show that when $H_e$ (respectively, $H_v$) satisfies the lower bound in~\eqref{eq:lb1_numOfPaths} (respectively,~\eqref{eq:lb2_numOfPaths}), there exists a feasible beam schedule that achieves the rate $\theta_c \widebar{\mathsf{C}}$. Moreover, we show that the lower bound in~\eqref{eq:lb1_numOfPaths} is also a necessary condition to ensure the target rate of $\theta_c \widebar{\mathsf{C}}$ when $M=1$.

\noindent$\bullet$ For $M=1$: Let $p_{[1:H_e]} \subseteq \mathcal{P}$ denote the \hbox{edge-disjoint} paths connecting the source to the destination. Since $M=1$ and we consider unit capacity links, we have $\widebar{\mathsf{C}} = 1$. We perform equal time sharing across the \hbox{edge-disjoint} paths in $p_{[1:H_e]}$ by operating each of them for a duration $\gamma \geq 0$ defined as
\begin{equation}
\label{eq:EdgeDisjOper}
\begin{aligned}
\gamma = \frac{\theta_c }{H_e} \widebar{\mathsf{C}}.
\end{aligned}
\end{equation}
We now show that the above is a feasible solution for the LP $\rm{P1}$ in~\eqref{capacity_paths}. 
In other words, we show that operating only $p_{[1:H_e]}$ (each for $\gamma$ of the time) provides a feasible solution for the LP $\rm{P1}$.
By the definition of $\gamma$ in~\eqref{eq:EdgeDisjOper}, the constraints in $(\rm{P1}a)$ are satisfied. For the constraints in $\rm{(P1}b)$ and $\rm{(P1}c)$, we note that $f^p_{p.nx(i),i} = 1$, $\forall i \in [0:N], \forall p \in p_{[1:H_e]}$ and $f^p_{i,p.pr(i)} = 1$, $\forall i \in [1:N+1], \forall p \in p_{[1:H_e]}$ because we consider unit link capacities. 
Thus, we can write $\rm{(P1}b)$ and $\rm{(P1}c)$ as
\begin{equation*}
\sum_{p \in p_{[1:H_e]}}\gamma = \sum_{p \in p_{[1:H_e]}}\frac{\theta_c }{H_e}\widebar{\mathsf{C}} = \theta_c \widebar{\mathsf{C}} \leq 1,
\end{equation*}
where the last inequality is due to the fact that $\theta_c \leq 1$ and $\widebar{\mathsf{C}} = 1$. This proves the feasibility of the solution for the LP $\rm{P1}$. 
Using this feasible solution, the achieved rate by the LP $\rm{P1}$ is given by $\theta_c \widebar{\mathsf{C}}$.
To satisfy the passive users constraint in~\eqref{passive_constraint}, we further set
\begin{equation}\label{eq:lemma_p1}
\begin{aligned}
\gamma = \frac{\theta_c }{H_e}\widebar{\mathsf{C}} \leq \theta \ \implies \ H_e \geq \frac{\theta_c}{\theta} \widebar{\mathsf{C}}.
\end{aligned}
\end{equation}
The above shows that, when $H_e$ satisfies~\eqref{eq:lemma_p1}, there exists a feasible beam schedule (i.e., time-sharing across $p_{[1:H_e]}$, each with duration $\gamma$) that achieves the rate $\theta_c \widebar{\mathsf{C}}$.

We now show that the condition on $H_e$ in~\eqref{eq:lemma_p1} is also necessary.
Since we consider unit capacity links, an optimal solution of this problem (i.e., maximizing the passive capacity) uses the paths in $p_{[1:H_e]}$, each with activation time $\theta$ due to the constraint in~\eqref{passive_constraint}. 
Thus, this solution leads to a passive capacity of $\mathsf{C} = \sum_{p \in p_{[1:H_e]}} \theta = H_e \theta$. This implies that, if we want to ensure a target rate of $\theta_c \bar{\mathsf{C}}$, we need that $H_e \theta \geq \theta_c \bar{\mathsf{C}}$, which implies that the condition on $H_e$ in~\eqref{eq:lemma_p1} is also necessary.


\smallskip
\noindent$\bullet$ For $M>1$: Let $p_{[1:H_v]}\subseteq \mathcal{P}$ be the \hbox{vertex-disjoint} paths connecting the source to the destination. Since $M>1$ and we consider unit capacity links, we have $\widebar{\mathsf{C}} = \min\left(M,H_v\right)$~\cite{Agarwal18}. We perform equal time sharing across the paths in $p_{[1:H_v]}$ by operating each of them for a duration $\gamma \geq 0$ defined as
\begin{equation}
\label{eq:VertexDisjOper}
\begin{aligned}
\gamma = \frac{\theta_c }{H_v} \widebar{\mathsf{C}}.
\end{aligned}
\end{equation}
We now show that the above choice satisfies the \hbox{1-2-1} constraints. 
Towards this end, we let $\mathcal{S}$ denote the set of states such that $|\mathcal{S}| = \max\left(M,H_v\right)$.
Each state $s \in \mathcal{S}$ activates $\min\left(M,H_v\right)$ of the paths in $p_{[1:H_v]}$ in a way that each $p \in p_{[1:H_v]}$ is activated in exactly $M$ number of states.
Thus, each state $s \in \mathcal{S}$ has an activation time equal to $\lambda_s = \gamma /M$. It therefore follows that each intermediate relay node in $p_{[1:H_v]}$ is used for transmission and reception for exactly $M$ times, each time for $\lambda_s$ of time. Thus, the amount of time that an intermediate relay node is used for transmission (similarly, for reception) is given by
\begin{equation*}
M \lambda_s = M \frac{\gamma}{M} = \gamma = \frac{\theta_c }{H_v} \widebar{\mathsf{C}} \leq 1,
\end{equation*}
where the inequality follows since $\theta_c \leq 1$ and $\widebar{\mathsf{C}}=\min\left(M,H_v\right) \leq H_v$. 
Thus, it is ensured that each intermediate relay node does not transmit (respectively, receive) for more than 100\% of the time.
Since the network is operated in $|\mathcal{S}| = \max\left(M,H_v\right)$ number of states and each state $s \in \mathcal{S}$ has an activation time equal to $\lambda_s = \gamma /M$, the amount of time each antenna at the source (respectively, destination) is used for transmission (respectively, reception) is given by
\begin{align*}
|\mathcal{S}| \lambda_s &= \max\left(M,H_v\right) \frac{\gamma}{M} = \max\left(M,H_v\right) \frac{1}{M} \frac{\theta_c }{H_v} \widebar{\mathsf{C}}
\\& = \max\left(M,H_v\right) \frac{1}{M} \frac{\theta_c }{H_v} \min\left(M,H_v\right) 
\\& = \frac{1}{M} \frac{\theta_c }{H_v} M H_v = \theta_c \leq 1.
\end{align*}
Thus, it is ensured that each antenna at the source (respectively, destination) is not used for transmission (respectively, reception) for more than 100\% of the time. This proves the feasibility of the proposed solution under \hbox{1-2-1} constraints. Using this feasible solution, the achieved rate is given by
\begin{equation*}
\min\left(M,H_v\right)  \lambda_s |\mathcal{S}| = \bar{\mathsf{C}} \frac{\gamma}{M} \max\left(M,H_v\right) = \theta_c \bar{\mathsf{C}}.
\end{equation*}
To satisfy the passive users constraint in~\eqref{passive_constraint}, we further set
\begin{equation}\label{eq:lemma_p2}
\begin{aligned}
\gamma = \frac{\theta_c }{H_v}\widebar{\mathsf{C}} \leq \theta \ \implies \ H_v \geq \frac{\theta_c}{\theta} \widebar{\mathsf{C}}.
\end{aligned}
\end{equation}
The above shows that, when $H_v$ satisfies~\eqref{eq:lemma_p2}, there exists a feasible beam schedule (i.e., time-sharing across $p_{[1:H_v]}$, each with duration $\gamma$) that achieves the rate $\theta_c \widebar{\mathsf{C}}$.
This concludes the proof of Theorem~\ref{lemma:numberOfPaths}.

\section{Proof of Theorem~\ref{lemma:security_connection1}}\label{appendix:security_connection1}
We start by proving the first direction, i.e., 
we start with an optimal solution for the passive user problem and we leverage it in the secure communication problem. We let $\mathcal{P}^\star$ denote the set of paths activated by an optimal passive user scheme (i.e., by the solution of the LP $\rm{P2}$ in~\eqref{capacity_lp} with the constraint in~\eqref{passive_constraint}) and $x_p^\star$ denote the corresponding activation times of the paths $p \in \mathcal{P}^\star$. 
In the secure communication problem, we use the same set of paths $\mathcal{P}^\star$ and the corresponding path activation times, which is a feasible solution for the secure communication problem since it satisfies the constraints in the LP $\rm{P2}$ in~\eqref{capacity_lp}. We assume that the eavesdropper wiretaps \textit{any} $K$ links in the network. Then, the secure capacity can be readily lower bounded by the following rate,
\begin{equation}\label{eq:passive_to_secure1}
R = \mathsf{C}-\max_{b \in \mathcal{B}}\sum_{(i,j) \in b}\theta_{ji},
\end{equation}
where $\mathsf{C}$ denotes the \hbox{1-2-1} passive capacity and $\mathcal{B}$ is the set of all combinations of $K$ links.
In particular,~\eqref{eq:passive_to_secure1} follows since
we consider unit capacity links and hence, every link $(i,j)$ that is wiretapped by the eavesdropper can cause at most a $\theta_{ji}$ reduction in the secure rate.
The maximization in~\eqref{eq:passive_to_secure1} ensures that we are considering the worst-case scenario.

We now focus on the second direction, i.e., we start with a secure communication scheme and we leverage it to create a feasible solution for the passive user problem such that it achieves the rate in~\eqref{eq:secure_to_passive}. In particular, depending on the value of $M$, we focus on two cases:

\noindent
{\bf{Case~1}:} $M=1$.
An optimal secure scheme~\cite{Agarwal18} uses the network for $H_e$ times and, at each instance, it activates
one of the paths from $p_{[1:H_e]}$, which denotes the \hbox{edge-disjoint} paths connecting the source to the destination. 
In other words, the optimal secure scheme in~\cite{Agarwal18} uses each path $p \in p_{[1:H_e]}$ for $1/H_e$ fraction of time to achieve the secure capacity of $\left(1-K/H_e\right)$. 
In the passive user problem, we can leverage the secure scheme of~\cite{Agarwal18} by using the same set of paths $p_{[1:H_e]}$.
In particular,
we select the activation time of each path $p \in p_{[1:H_e]}$ as $\min\left(1/H_e,\theta_p\right)$, where $\theta_p {=} \min_{(i,j) \in \mathcal{E}_p}\theta_{ji}$.
We note that this choice: (i) achieves the rate in~\eqref{eq:secure_to_passive}; and (ii) satisfies the \hbox{1-2-1} constraints in the LP $\rm{P1}$ in~\eqref{capacity_paths}. In particular, the constraints in $(\rm{P1}a)$ are satisfied since $\min\left(1/H_e,\theta_p\right) {\geq} 0$. For the constraints in $\rm{(P1}b)$ and $\rm{(P1}c)$, since we consider unit link capacities, we have $f^p_{p.nx(i),i} = 1$, $\forall i \in [0:N], \forall p \in p_{[1:H_e]}$ and $f^p_{i,p.pr(i)} = 1$, $\forall i \in [1:N+1], \forall p \in p_{[1:H_e]}$. Thus, we can write $\rm{(P1}b)$ and $\rm{(P1}c)$ as
\begin{equation*}
\begin{aligned}
\sum_{p \in p_{[1:H_e]}}  \min\left(\frac{1}{H_e},\theta_p\right) \leq \sum_{p \in p_{[1:H_e]}} \frac{1}{H_e} = 1.
\end{aligned}
\end{equation*}

Moreover, for every path $p {\in} p_{[1:H_e]}$, we have that $\min\left(1/H_e,\theta_p\right) {\leq} \theta_p {\leq} \theta_{ji}$ for all $(i,j)  \in \mathcal{E}_p$. Thus, the passive users constraint in~\eqref{passive_constraint} is also satisfied.

\noindent
{\bf{Case~2}:} $M>1$. The secure scheme proposed in~\cite{Agarwal18} uses paths in $p_{[1:H_v]}$ for transmission. Since the intermediate relay nodes can transmit to only one node and receive from only one node, $\hat{M} = \min\left(M,H_v\right)$ paths can be simultaneously active at each time the network is used. The secure scheme uses the network for $\binom{H_v}{\hat{M}}$ times and, at each instance, a different choice of $\hat{M}$ paths is used. Particularly, every path $p \in p_{[1:H_v]}$ is used for $\binom{H_v-1}{\hat{M}-1}$ times. Since an equal time is allocated to every network usage, each path has an activation time:
\begin{equation*}
\binom{H_v-1}{\hat{M}-1}\frac{1}{\binom{H_v}{\hat{M}}} = \frac{\hat{M}}{H_v}.
\end{equation*}
We note that the \hbox{1-2-1} constraints for every node in the network are satisfied since $\hat{M}/H_v \leq 1$ and hence, none of the antennas is used for more than 100\% of the time.
In the passive user scheme, we use the same set of paths $p_{[1:H_v]}$, and we set the activation time of each path $p \in p_{[1:H_v]}$ equal to $\min\left(\hat{M}/H_v, \theta_p\right)$ to ensure that the constraint in~\eqref{passive_constraint} is satisfied.
With this, the rate that can be guaranteed in the passive user problem is given by,
\begin{equation*}
R = \sum_{p \in p_{[1:H_v]}}\min\left(\frac{\hat{M}}{H_v},\theta_p\right)= \sum_{p \in p_{[1:H_v]}}\min\left(\frac{M}{H_v},\theta_p\right),
\end{equation*}
where the last equality follows since $\theta_p \leq 1$.
This concludes the proof of Theorem~\ref{lemma:security_connection1}.

\end{document}